\documentclass{aa}  

\usepackage{graphicx}
\usepackage{txfonts}
\begin{document}
\title{WD0433$+$270: an old Hyades stream member or an Fe-core white 
dwarf?$^\star$}

\author{S. Catal\'an
       \inst{1,2},
       I. Ribas
       \inst{1,2}
       J. Isern
       \inst{1,2}
           \and 
       E. Garc\'\i a--Berro
       \inst{1,3}
}

\offprints{S. Catal\'an \\
$^\star$Based  on observations  obtained at:  Calar  Alto Observatory,
Almer\'\i a, Spain and McDonald Observatory, Texas, USA.}

\institute{Institut d'Estudis Espacials de Catalunya, c/ Gran Capit\`{a} 
           2--4, 08034 Barcelona, Spain\
  	   \and	     
           Institut de Ci\`encies de l'Espai, CSIC, Facultat  de
           Ci\`encies, Campus UAB, 08193 Bellaterra, Spain\           
	   \and	     
	   Departament de F\'isica Aplicada, 
           Escola Polit\`ecnica Superior de Castelldefels, 
           Universitat Polit\`ecnica de Catalunya, 
           Avda. del Canal Ol\'\i mpic s/n, 08860 
	   Castelldefels, Spain}
\date{\today}
\abstract
{G39$-$27/289 is a  common proper motion pair formed  by a white dwarf
(WD0433$+$270) and a main-sequence  star (BD$+$26~730) that apparently
has been classified  as a member of the  Hyades open cluster. Previous
studies of the white dwarf component yielded a cooling time of $\sim4$
Gyr.  Although it  has not  been pointed  out before  explicitly, this
result is  6 times larger than  the age of the  Hyades cluster, giving
rise to an  apparent conflict between the physics  of white dwarfs and
cluster main-sequence fitting.}
{We investigate whether this system belongs to the Hyades cluster and,
accordingly, give a  plausible explanation to the nature  of the white
dwarf member.}
{We have  performed and analyzed spectroscopic  observations to better
characterize  these objects,  and used  their kinematic  properties to
evaluate their membership to  the Hyades.  Then, different mass-radius
relations and  cooling sequences for different  core compositions (He,
C/O, O/Ne  and Fe) have  been employed to  infer the mass  and cooling
time of the white dwarf.}
{From  kinematic and  chemical composition  considerations  we believe
that  the  system  was a  former  member  of  the Hyades  cluster  and
therefore has an  evolutionary link with it. However,  the evidence is
not  conclusive.  With  regards  to  the nature  of  the  white  dwarf
component, we find  that two core compositions --- C/O  and Fe --- are
compatible with  the observed effective temperature  and radius. These
compositions  yield very different  cooling times  of $\sim$4  Gyr and
$\sim$1 Gyr, respectively.}
{We distinguish two  posssible scenarios. If the pair  does not belong
to  the Hyades  cluster  but only  to  the Hyades  stream, this  would
indicate that  such stream contains rather old  objects and definitely
not coeval  with the cluster.   This has interesting  consequences for
Galactic dynamics. However,  our favoured scenario is that  of a white
dwarf with a  rather exotic Fe core, having  a cooling time compatible
with the  Hyades age.   This is a  tantalizing result that  would have
implications  for  the thermonuclear  explosion  of  white dwarfs  and
explosion theories of degenerate nuclei.}

\authorrunning{Catal\'an et al.} 
\titlerunning{WD0433$+$270:  an   old  Hyades  stream   member  or  an
              Fe-core white dwarf?}  
\maketitle

%_____________________________________________________________________

\section{Introduction}

The determination  of reliable ages  is of obvious importance  to both
astrophysics and  cosmology, but not exempt of  many complications ---
see,  e.g., \cite{von01}.   The  age of  a  star is  perhaps the  most
difficult property to estimate  and, further, it nearly always depends
on rather strong model assumptions (Mamajek et al.~2007). An exception
is the method that uses radioactive decay to directly estimate stellar
ages (Cayrel  et al.~2001;  Frebel et al.~2007),  but this  has rather
restricted    applicability.     Another    emerging   technique    is
asteroseismology,  which has  the potential  to provide  accurate ages
(Miglio  \& Montalb\'an  2005), but  this will  require high-precision
photometric  data from  upcoming space  missions and  also  stars with
well-constrained physical properties. In the meantime, the use of open
clusters and main sequence stellar evolutionary models continues to be
the most widely used method to  infer the ages of stars in the Galaxy.
In  spite  of  the  still-present uncertainties,  such  as  convective
overshoot, chemical composition anomalies,\ldots~the field has reached
sufficient maturity to  provide ages that are reliable  to better than
$\sim$10\% --- see, for  instance, \cite{pau06}.  The analogous method
of  using  eclipsing  binaries  (Ribas  et  al.~2000)  yields  similar
(model-dependent) accuracy.

The  study of  white dwarfs  has made  very valuable  contributions to
numerous  areas of  astrophysics,  and estimating  individual ages  of
stars  is no  exception. The  main advantage  of white  dwarfs  is the
conceptual simplicity of their evolution,  which can be described as a
simple  cooling process.   Modelling  of the  cooling sequences  makes
white  dwarfs powerful  stellar chronometers,  accurate to  about 25\%
(Silvestri et al.~2005).  Among the major uncertainties of this method
are the pre-white dwarf evolution time and the effects of the chemical
composition of the core.

\begin{table*}[t]
\begin{center}
\caption{Photometric data and stellar parameters derived for BD$+$26~730.}
\small{
\begin{tabular}{lccccccc}
\hline
\hline
\noalign{\smallskip}
$V$ & $J$ & $H$ & $K$  & $T_{\rm eff}$ (K) &  [Fe/H] & $Z$ & $\log(L/L_{\sun})$ \\
\noalign{\smallskip}
\hline
\noalign{\smallskip}
8.42$\pm$0.02 & 5.945$\pm$0.023 & 5.400$\pm$0.018 & 5.240$\pm$0.023 & $4,595\pm30$ & $0.03\pm0.09$   & $0.021\pm0.004$  & $-0.527\pm0.021$ \\ 
\noalign{\smallskip}
\hline
\hline
\end{tabular}
\label{tab:spar}}
\end{center}
\end{table*}

In this paper we discuss a case that presents an interesting puzzle in
which the age estimate stemming  from the white dwarf cooling sequence
is  in apparent  conflict with  the age  estimate coming  from cluster
membership. The  object is G39$-$27/289,  a common proper  motion pair
formed  by a  DA white  dwarf  (WD0433$+$270) and  a K  type star  ---
BD$+$26~730 (Holberg et al.~2002) --- which is a well studied variable
star (V833  Tau) and also  a single-lined spectroscopic binary  with a
very  low-mass companion (Tokovinin  et al.~2006).   The members  of a
common proper motion pair were likely born simultaneously and with the
same chemical composition (Wegner 1973; Oswalt et al.~1988). Since the
components are well separated ($\sim  2,200$ AU in this case), no mass
exchange  has taken  place and  they have  evolved as  isolated stars.
Thus, it  is logical  to assume that  both components  of G39$-$27/289
have  the same  age  and  the same  original  metallicity. The  K-star
component, and, by extension, its common proper motion companion, were
classified as Hyades members  by \cite{per98} and therefore would have
an age  ranging from about  0.6 to 0.7  Gyr.  However, this  stands in
obvious conflict with the cooling time of WD0433$+$270, which has been
estimated to be about 4 Gyr (Bergeron et al.~2001).

In this  work, we carry out  a detailed study of  these objects, using
both  information   present  in  the  literature  and   also  our  own
spectroscopic  observations,  with the  objective  of unveiling  their
nature and their  possible membership to the Hyades  open cluster.  We
evaluate  the   different  scenarios  and   discuss  their  respective
implications to white dwarf physics and Galactic dynamics.

\section{Observations} 
\label{sec:obs}

We have  observed both  components of the  common proper  motion pair,
namely   BD$+$26~730   and  WD0433$+$270,   with   the  objective   of
characterizing their radiative properties. In the case of BD$+$26~730,
we  performed  spectroscopic   observations  with  the  FOCES  echelle
spectrograph  on the  2.2~m telescope  at CAHA  (Almer\'{\i}a, Spain),
obtaining  a resolution  of  $R\sim 47,000$.   The  data were  reduced
following the standard procedures using  the echelle tasks of the IRAF
package.  From  these observations, together with a  detailed study of
the  visible and  near-infrared  spectral energy  distribution of  the
star,  we  derived  the   effective  temperature  and  metallicity  of
BD$+$26~730  following the procedure  described in  \cite{cat07}.  The
resulting  value of $T_{\rm  eff}=4,595\pm30$ K  is in  good agreement
with  the  determination of  \cite{ola01}.   The chemical  composition
analysis     yielded     a     nearly     solar     metallicity     of
[Fe/H]$=0.03\pm0.09$. The  total luminosity, $L$,  was calculated from
the  apparent  magnitude  and   a  bolometric  correction  (Masana  et
al.~2006).   A summary  of  the   photometric  information  of
BD$+$26~730  and the derived  stellar  parameters  is  given in  Table
\ref{tab:spar}.

A   very  relevant   parameter   for   our  study   is   the  age   of
BD$+$26~730.  Unfortunately, this cannot  be determined  reliably from
the observed $T_{\rm eff}$ and  $L$ and using model isochrones because
of the relative  proximity of the star to  the zero-age main sequence.
Note, however,  that \cite{bar96} determined lithium  abundances for a
sample   of   binaries  classified   as   Hyades  members,   including
BD$+$26~730. Lithium was  indeed detected in this case, obtaining
a value  of $\log N_{\rm Li}=0.31$,  which is in good  accord with the
rest  of the  observed  binaries, albeit  with  an abundance  somewhat
larger  than that  of single  objects belonging  to the  Hyades ($\log
N_{\rm  Li}=-0.5$).   According  to  \cite{bar96}, this  is  something
expected since,  in general, binary  systems have an  overabundance of
lithium   with   respect  to   single   stars.   Later,   \cite{bar97}
recalculated the  lithium abundance of BD$+$26~730,  obtaining in this
case $\log N_{\rm  Li}=0.23$.  They compared the Li  abundance and the
effective   temperature  of   this  object   with   lithium  depletion
isochrones,  yielding an  age of  $\sim$  600 Myr.  Thus, the  results
suggest a Hyades age for BD$+$26~730.

In  the case  of WD0433$+$270,  a cool  white dwarf,  the best  way to
estimate its effective temperature  is by using the photometric energy
distribution   instead   of   a   spectroscopic   fit   (Bergeron   et
al.~2001). However,  we decided to  observe this object to  ensure its
spectral classification. The observations  were performed with the LCS
spectrograph  of the  Harlan  J.~Smith (2.7~m)  telescope at  McDonald
Observatory (Texas, USA) covering some  of the main Balmer lines (from
3,885 to 5,267~\AA)  and obtaining a resolving power  of $\sim 5$~\AA\
FWHM.  We used the standard procedures within the single-slit tasks of
the IRAF package. The spectrum shows weak absorption lines but we were
able to unambiguously identify H$\beta$.  The weakness of the lines is
the reason why  this star had been previously classified  as a type DC
white dwarf  by several authors  (Eggen \& Greenstein 1965;  Oswalt \&
Strunk  1994).   Higher resolution  observations  by \cite{ber01}  and
\cite{zuc03}  had  permitted   the  identification  of  H$\alpha$  and
H$\beta$,   respectively.   Putting  all   results  together   we  can
confidently classify this object as a DA white dwarf.

\begin{table}[t]
\begin{center}
\caption{Photometric  properties  and  parallax of  WD0433$+$270.  The
         uncertainties are typically 3\% for $V$, $R$, and $I$, and 5\% 
         for the rest.}  
\small{
\begin{tabular}{lcccccccc}
\hline
\hline
\noalign{\smallskip}
$B$ & $V$ & $R$ & $I$ & $J$ & $H$ & $K$ & $\pi$ (mas)\\
\noalign{\smallskip}
\hline
\noalign{\smallskip}
16.48 & 15.81 & 15.40 & 15.01 & 14.61 & 14.42 & 14.32  & 60$\pm$3\\
\noalign{\smallskip}
\hline
\hline
\end{tabular}
\label{tab:wdphot}}
\end{center}
\end{table}

\begin{table}
\begin{center}
\caption{Atmospheric  parameters  of  WD0433$+$270  available  in  the
         literature.}
\begin{tabular}{lccc}
\hline
\hline
\noalign{\smallskip}
$T_{\rm eff}$ (K)& $\log g$ &  Reference \\
\noalign{\smallskip}
\hline
\noalign{\smallskip}
 $5,620\pm110$  &$8.14\pm0.07$ & \cite{ber01} \\
 $5,434\pm300$  &$8.0\pm0.1$   & \cite{zuc03} \\
\noalign{\smallskip}
\hline
\hline
\end{tabular}
\label{tab:wd2}
\end{center}
\end{table}

\begin{table*}[t]
\begin{center}
\caption{Spacial  velocities  for the  studied  stars  and the  Hyades
         cluster and streams.}
\begin{tabular}{lccccccc}
\hline
\hline
\noalign{\smallskip}
 & $\mu_{\alpha}$ & $\mu_{\delta}$ &$V_r$ & $U\pm\sigma_U$ & $V\pm\sigma_V$ & $W\pm\sigma_W$ \\
 & (mas/yr) & (mas/yr) & (km/s) & (km/s) & (km/s) & (km/s) \\
\noalign{\smallskip}
\hline
\noalign{\smallskip}
WD0433$+$270$^1$ & 228    & $-155$    & $+36.3$         & $-39.2$        & $-15.5$       & $-1.8$       \\
WD0433$+$270$^2$ & 228    & $-155$    & $+41.7$         & $-44.4$        & $-15.7$       & $-3.9$       \\
BD$+$26~730$^3$  & 232.36 & $-147.11$ & $+36.18\pm0.08$ & $-39.4$        & $-17.2$       & $-1.6$       \\
Hyades OCl$^4$	 & $...$  & $...$     & $...$           & $-42.8\pm3.6$  & $-17.9\pm3.2$ & $-2.2\pm5.2$ \\
Hyades SCl(1)$^4$& $...$  & $...$     & $...$           & $-31.6\pm2.8$  & $-15.8\pm2.8$ & $ 0.8\pm2.7$ \\
Hyades SCl(2)$^4$& $...$  & $...$     & $...$           & $-33.0\pm4.2$  & $-14.1\pm4.0$ & $-5.1\pm3.1$ \\
Hyades SCl(3)$^4$& $...$  & $...$     & $...$           & $-32.8\pm2.8$  & $-11.8\pm2.8$ & $-8.9\pm2.9$ \\
Hyades Stream$^5$& $...$  & $...$     & $...$           & $-30.3\pm1.5$  & $-20.3\pm0.6$ & $-8.8\pm4.0$ \\
\noalign{\smallskip}
\hline
\hline
\end{tabular}
\label{tab:vel}
\end{center}
\small \footnotemark[1]{Proper motions from \cite{sal03} and velocities from \cite{zuc03}.}\\
\small \footnotemark[2]{Proper motions from \cite{sal03} and velocities recalculated considering a Fe core (see \S4).}\\
\small \footnotemark[3]{Spacial velocities calculated from the radial velocity and proper motions reported by \cite{per97}.}\\
\small \footnotemark[4]{Spacial velocities from \cite{che99}.}\\
\small \footnotemark[5]{Spacial velocities from \cite{fam05}.}\\
\end{table*}

\cite{ber01}   used   $BVRIJHK$   photometry  ---   see   Table
\ref{tab:wdphot}  ---  to  obtain  the energy  flux  distribution  of
WD0433$+$270.    From   a   fit   of  different   theoretical   energy
distributions they  obtained its effective  temperature and $(R/d)^2$,
where $R$  is the  radius of the  star and  $d$ is the  distance.  The
radius can  be determined accurately because  a trigonometric distance
is available for  this object. The authors also  obtained its mass and
$\log g$ from  the mass-radius relation of \cite{fon01}  for C/O white
dwarfs.  \cite{zuc03} followed the same procedure as \cite{ber01}, but
using the mass-radius relation  of \cite{woo95}.  The results obtained
by these authors  are summarized in Table \ref{tab:wd2}.   It is worth
mentioning that  the spectroscopic observations  were not used  in any
case to estimate the stellar properties but they served as an internal
check of the photometric solutions via comparison of calculated models
with the observed line profiles.

The luminosity  of the white dwarf  can be derived  from its effective
temperature (Bergeron et al.~2001) and distance ($d=16.95\pm0.86$ pc),
and considering its apparent  magnitude and the bolometric corrections
of  \cite{mas06}.  The  value  that we  obtained is  $\log(L/L_{\sun})
=-3.92\pm0.04$,   and,   from   this,   a   radius   of   $R=0.0115\pm
0.0010~R_{\sun}$.   Both  results are  in  good  agreement with  those
reported by \cite{ber01}.

\section{Membership to the Hyades}

The membership of WD0433$+$270 to  the Hyades cluster was evaluated by
\cite{egg93a} within  the course  of a study  of all  degenerate stars
with certain  kinematic criteria.  WD0433$+$270 was  excluded from the
member  list,  although  the  radial  velocity  of  its  companion  is
identical to  the value predicted from  membership.  The justification
was a color index supposedly too red to belong to the Hyades but, more
importantly,  the distance  of the  common proper  motion pair  to the
center of the cluster.  This  author concluded that this common proper
motion pair is  projected on the cluster but only at  about a third of
the distance to the cluster.

The comprehensive study of \cite{per98} considered the 5,490 Hipparcos
Catalog  stars corresponding  to the  field of  the  Hyades. Hipparcos
astrometry was combined with  radial velocity measurements in order to
obtain   three-dimensional   velocities,   which   allowed   candidate
membership  selection based  on position  and kinematic  criteria. The
authors  divided  the  Hyades   into  four  components  by  using  the
three-dimensional  distance to  the cluster  center.  The  distance of
BD$+$26~730 to the center of the cluster is 29 pc and therefore it was
classified as a  former member of the Hyades  cluster, currently lying
beyond  the tidal  radius  ($\sim$10 pc).   In contrast,  \cite{deb99}
preferred not  to include BD$+$26~730 in their  list of  Hyades member
stars   following  the   conclusions   of  a   study   based  on   the
convergent-point method.

It is also  worth mentioning two further pieces  of evidence that were
not included in the studies  mentioned above. One is the metal content
of BD$+$26~730. As discussed in Sect.  \ref{sec:obs}, we have reliably
determined  the metallicity  of  this  star and  obtained  a value  of
[Fe/H]$=0.03\pm0.09$. This  is in  reasonably good agreement  with the
Hyades   metallicity   of   [Fe/H]$=0.14\pm0.05$  as   determined   by
\cite{per98}.  The result is not conclusive, however, because the mean
metallicity of field stars is of [Fe/H]$=-0.14\pm0.19$ (Nordstr\"om et
al.~2004). The  other important point  is the detection of  lithium in
the spectrum  of BD$+$26~730 by  \cite{bar96}, which clearly favours a
relatively young age  for this object, and in  agreement with the rest
of the Hyades members studied.

It has been often mentioned that there is a spatially unbound group of
stars  in the  solar neighbourhood  with  the same  kinematics as  the
Hyades open  cluster (Eggen et al.~1993b; Perryman  et al.~1998). This
group   of   stars   is   called   the   Hyades   stream   or   Hyades
supercluster. \cite{che99} mapped the density-velocity inhomogeneities
of an  absolute magnitude  limited sample of  A--F type  dwarfs. Three
different clumps within the Hyades stream were distinguished, each one
of them with characteristic space velocities, which are given in Table
\ref{tab:vel}.   The  authors  also  claimed that  the  Hyades  stream
contains probably three groups of  0.5--0.6 Gyr, 1 Gyr and 1.6--2 Gyr,
which  are in an  advanced stage  of dispersion  in the  same velocity
volume.   Each  stream  presents  a characteristic  age  distribution,
although  the  velocity  separation  does  not  produce  a  clear  age
separation.

In Table \ref{tab:vel}  we provide the radial and  space velocities of
BD$+$26~730 and WD0433$+$270. We also give the kinematic properties of
the  Hyades open cluster  (OCl) and  of each  clump within  the stream
according  to   \cite{che99}  (called  SCl   1,  2  and  3   by  these
authors).  The   recent  spatial  velocities  of   the  Hyades  stream
calculated  by \cite{fam05}  are also  listed.  As  can be  seen, both
members of  the common proper  motion pair have  velocities compatible
with those of the Hyades  open cluster and are somewhat different from
the  velocities characteristic  of  the Hyades  stream  or the  clumps
within the stream.

From the kinematic considerations made here, together with the lithium
detection  in BD$+$26~730, we favour  the hypothesis  that  the common
proper  motion pair  studied here  is  indeed linked  with the  Hyades
cluster evolution-wise. It is certainly  not a {\sl bona fide} cluster
member because of  it location beyond the tidal  radius of the cluster
but it is likely a former member that has escaped. If this scenario is
correct, the components of the pair  should have the age of the Hyades
cluster,   which    was   estimated   to   be    $625\pm50$   Myr   by
\cite{per98}. However,  we do not have  conclusive evidence supporting
this  evolutionary link  and therefore  the alternative  scenario that
this pair belongs to the Hyades stream cannot be completely ruled out.

%_____________________________________________________________________

\section{The nature of WD0433$+$270}

\begin{figure}[t]
\begin{center}
\includegraphics[clip,width=1.0\columnwidth]{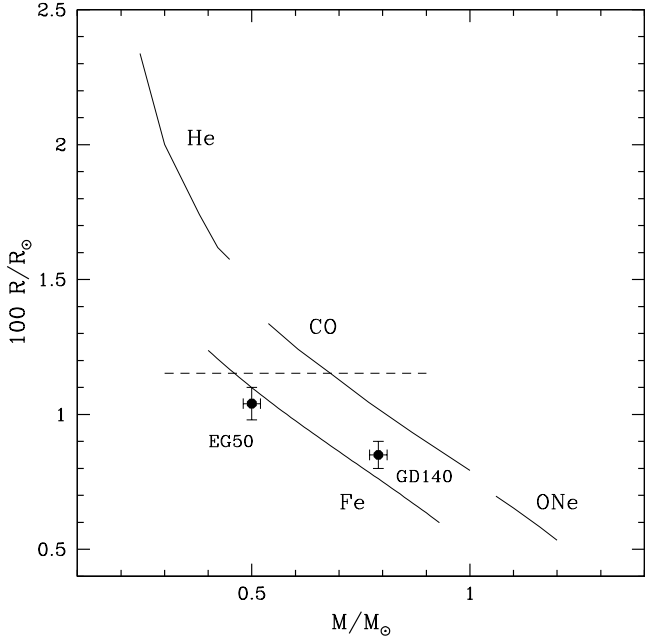}
\caption{Mass-radius relations for  different compositions and $T_{\rm
         eff}=5,500$ K. The dashed line  corresponds to the  radius of
         WD0433$+$270.  The filled circles correspond to observational
         data obtained by \cite{pro98}.}
\label{fig:massrad}
\end{center}
\end{figure}

Most white dwarfs have a core made of C/O, although other compositions
are  possible.   However,  in  this  particular case,  and  given  the
putative age discrepancy with the  Hyades cluster, we consider all the
feasible   compositions  that   have  been   proposed  to   date.   In
Fig.~\ref{fig:massrad}  we   show  mass-radius  relations  considering
different compositions for the core:  He  (Serenelli et  al.~2002),  C/O
(Salaris et  al.~2000), O/Ne  (Althaus et al.~2005)  and Fe  (Panei et
al.~2000a).    All  these   relations  correspond   to   an  effective
temperature of $5,500$ K.  As  can be seen, the radius of WD0433$+$270
only has a corresponding mass in  the case of C/O and Fe cores (dashed
line).  The existence  of Fe white dwarfs has been  pointed out in the
past.  \cite{pro98} derived the radii of some field white dwarfs using
parallaxes and effective temperatures.  Then, they obtained the masses
from the radii and surface gravities, which were previously determined
independently from  spectroscopic fits to white dwarf  models.  In two
cases (GD~140 and  EG~50) they obtained masses that  were too small to
fit  the C/O  mass-radius relation.   The best  explanation  for these
stars that the  authors proposed is that they have  a Fe core (Shipman
\& Provencal 1999).   The data corresponding to these  stars have been
also plotted in Fig.~\ref{fig:massrad}.

\begin{table}
\begin{center}
\caption{Masses and cooling times  for WD0433$+$270 assuming a C/O and
         a Fe core.}
\begin{tabular}{ccc}
\hline
\hline
\noalign{\smallskip}
Model & $M_{\rm WD} \,(M_{\sun})$ & $t_{\rm cool}$~(Gyr) \\
\noalign{\smallskip}
\hline
\noalign{\smallskip}
C/O core & $0.67\pm0.03$ & $4.1\pm1.2$ \\
Fe core  & $0.46\pm0.07$ & $1.0\pm0.1$ \\
\noalign{\smallskip}
\hline
\hline
\end{tabular}
\end{center}
\label{tab:mascol}
\end{table}

The gravitational  redshift and the  radial velocity of a  white dwarf
are two interdependent parameters.  \cite{zuc03} calculated the radial
velocity  of WD0433$+$270  (see  Table \ref{tab:vel})  assuming a  C/O
core.   Since a  Fe core  seems also  possible in  this case,  we have
recalculated  this  value  considering  the  mass-radius  relation  of
\cite{pan00b} for Fe white dwarfs.  As seen in Table \ref{tab:vel}, in
both  cases,  radial and  space  velocities  are  compatible with  the
kinematic properties  of the  Hyades cluster, considering  the typical
errors of white dwarf radial velocities (Schultz et al.~1996).

Turning  the argument  around,  the  kinematics can  also  be used  to
estimate the mass of  WD0433$+$270 independently of the composition of
the core.  This can  be done assuming  that WD0433$+$270 has  the same
radial  velocity as  its companion,  BD$+$26~730, and  considering the
observed velocity  of this white dwarf reported  by \cite{zuc03}. This
yields a  mass of $\sim0.55$  $M_{\sun}$, which does  not conclusively
favour any of the two core compositions (C/O or Fe).

\begin{figure*}[t]
\begin{center}
\includegraphics[clip,width=0.8\columnwidth]{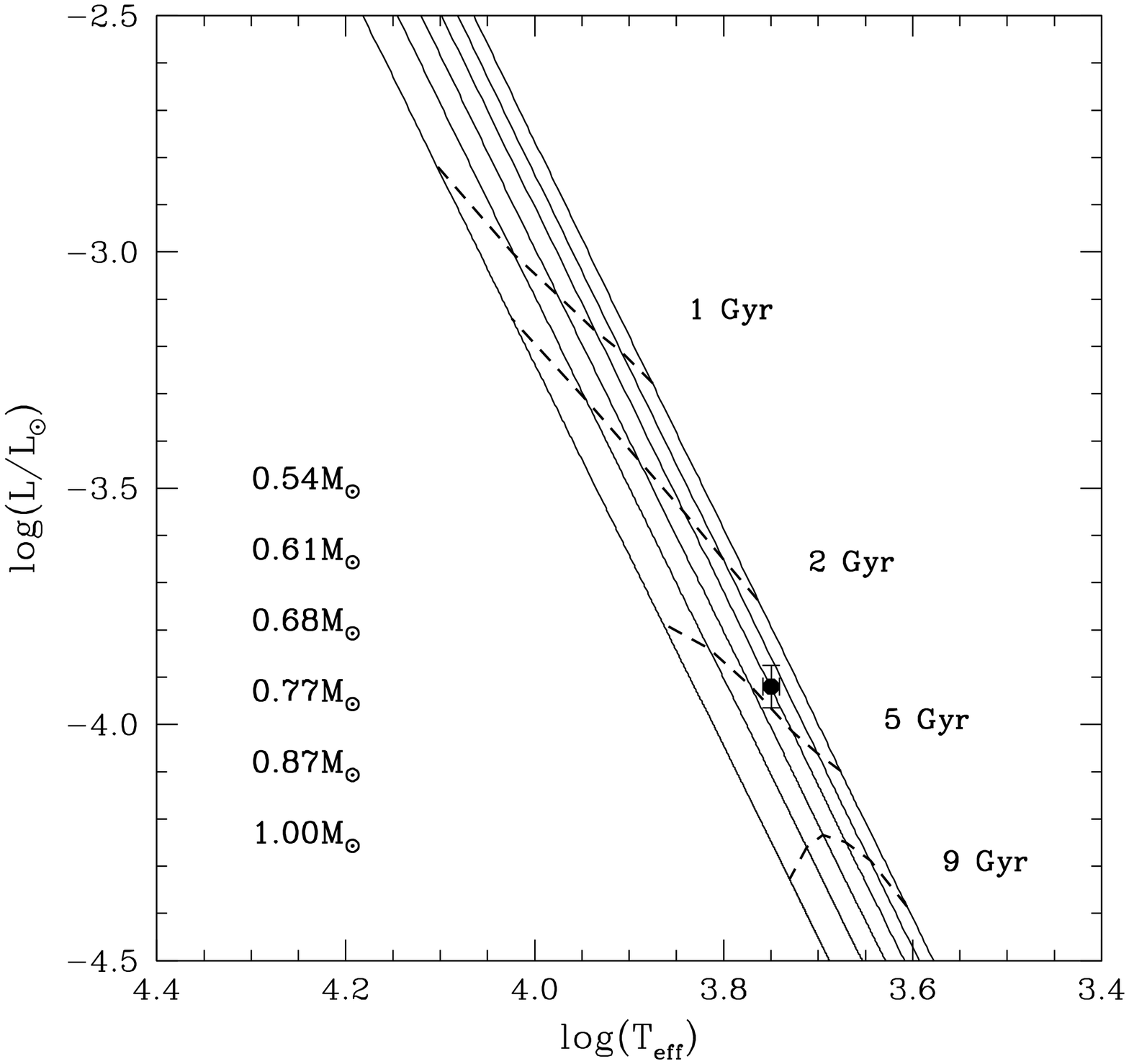}
\includegraphics[clip,width=0.8\columnwidth]{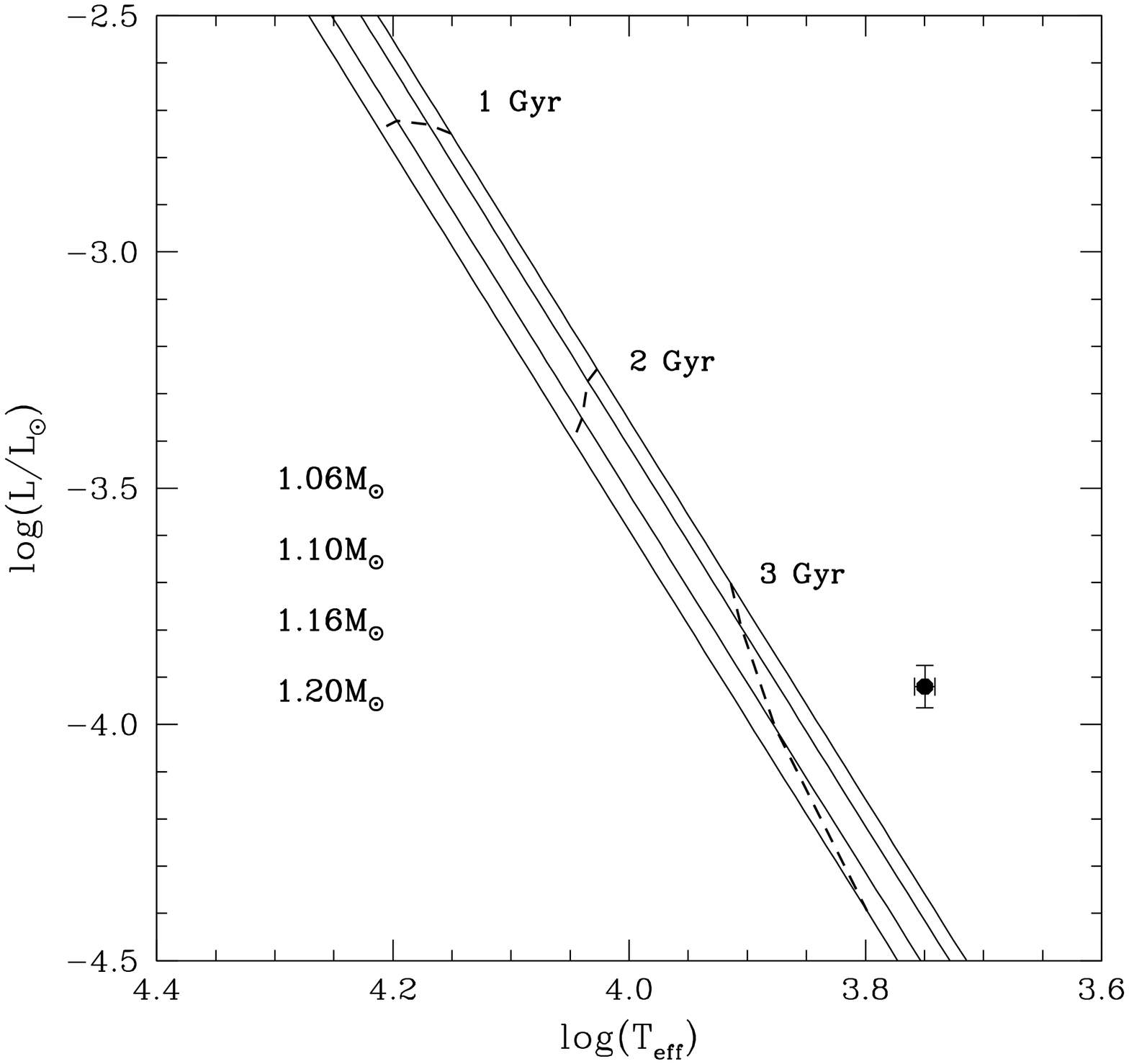}
\includegraphics[clip,width=0.8\columnwidth]{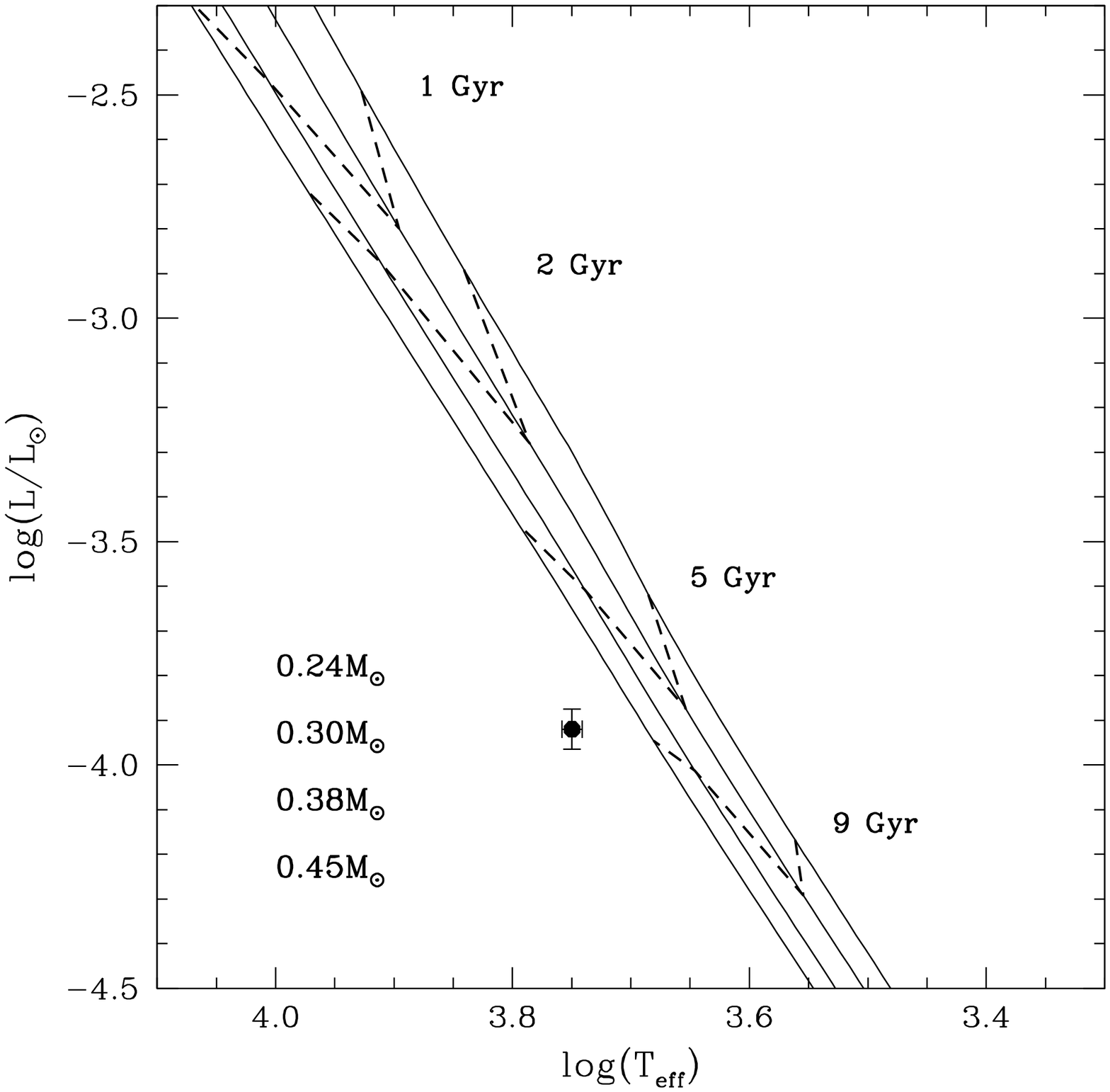}
\includegraphics[clip,width=0.8\columnwidth]{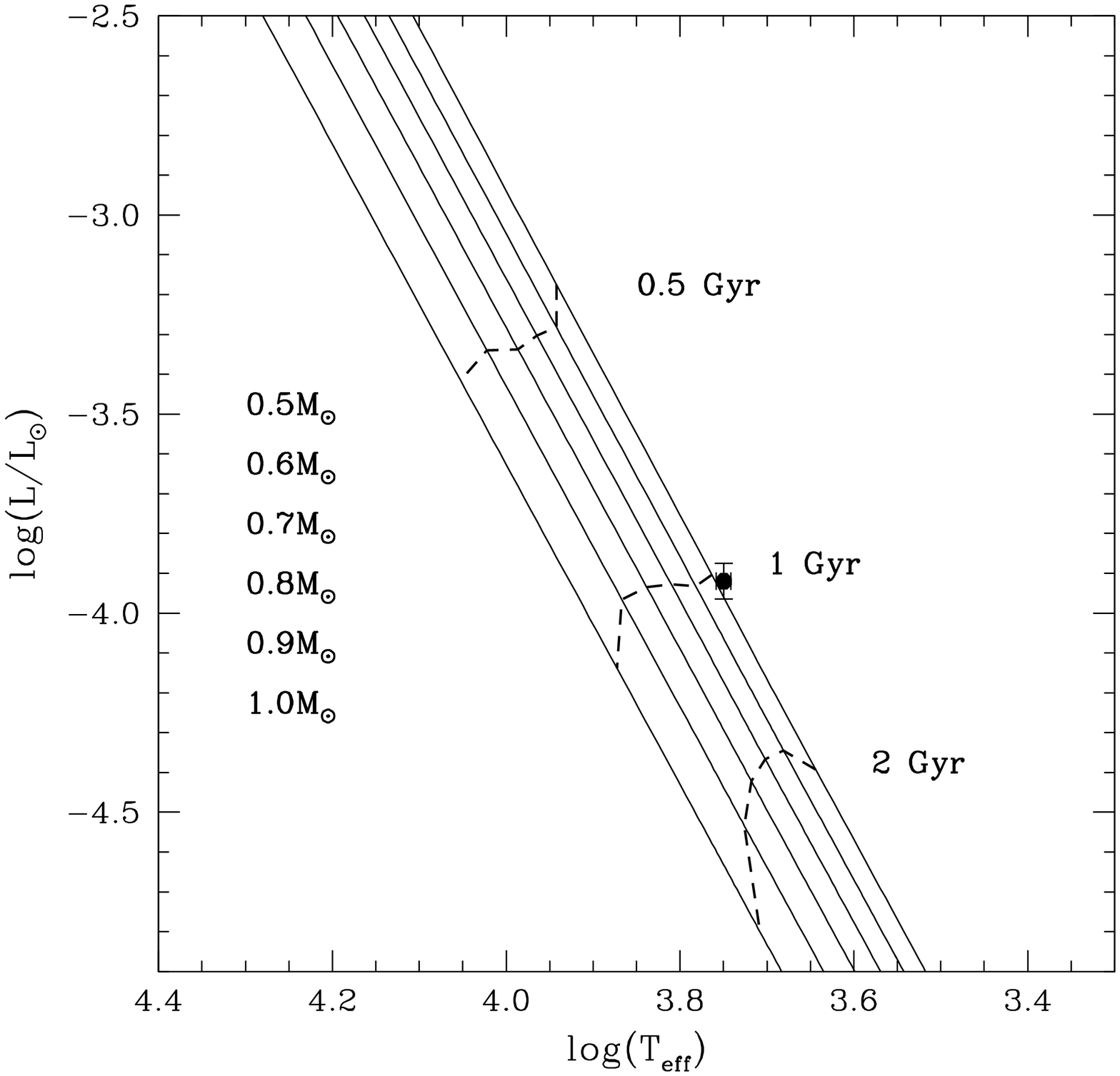}
\caption{Luminosity versus effective temperature for different cooling
         tracks:  C/O-core (top-left), O/Ne-core  (top-right), He-core
         (bottom-left)  and Fe-core  (bottom-right). The  dashed lines
         represent cooling isochrones.}
\label{fig:lumtef}
\end{center}
\end{figure*}

Once we know  the masses that correspond to  each mass-radius relation
(see Table \ref{tab:mascol}) we  can calculate the cooling times using
the  proper cooling  sequences. In  Fig.~\ref{fig:lumtef} we  show the
cooling  sequences for  different core  compositions:  C/O (top-left),
O/Ne (top-right),  He (bottom-left) and Fe  (bottom-right). The dashed
lines  correspond  to cooling  isochrones.   We  have overplotted  the
effective temperature and luminosity  of WD0433$+$270 to check whether
it  falls within  the range  of values  corresponding to  each  set of
cooling sequences.  As expected, there  is no correspondence  with the
O/Ne- and  He-core model sequences  since the physics of  white dwarfs
with these compositions  predicts masses for this radius  that are too
high or too low, respectively.

The most common internal structure of a white dwarf is thought to be a
core  made of  C/O surrounded  by a  H thick  envelope ontop  of  a He
buffer,  with  compositions of  $q({\rm  H})=M_{\rm H}/M=10^{-4}$  and
$q({\rm   He})=M_{\rm  He}/M=10^{-2}$.    The  cooling   sequences  of
\cite{sal00} consider such configuration,  but with a larger abundance
of  O than  C  at the  center  of the  core.   These improved  cooling
sequences include an accurate treatment of the crystallization process
of  the C/O  core,  including phase  separation upon  crystallization,
together  with state-of-the-art input  physics suitable  for computing
white dwarf evolution.  As can  be seen in Table \ref{tab:mascol}, the
cooling time derived ($\sim$4 Gyr)  is in excellent agreement with the
results obtained by \cite{ber01}. This  cooling time is 6 times larger
than the age  of the Hyades open cluster and 2  times larger than that
of the  older Hyades  stream group ---  about $\sim$2 Gyr  (Chereul et
al.~1999).  The assumption  of  an O/Ne-core  white  dwarf would  give
smaller cooling times, since the heat  capacity of O and Ne is smaller
than that  of C, and as result  white dwarfs of this  type cool faster
(by a factor  of 2). We considered the  cooling tracks of \cite{alt07}
for  O/Ne-core white  dwarfs (Fig.~\ref{fig:lumtef},  top-right).  The
outer  layer  chemical  stratification  consists of  a  pure  hydrogen
envelope  of $10^{-6}M_{\sun}$ overlying  a helium-dominated  shell of
$4\times10^{-4}M_{\sun}$ and,  below that,  a buffer rich  in $^{12}$C
and $^{16}$O. However, as mentioned before, this core composition does
not reproduce a white dwarf with such small radius.

In the light of the recent results obtained by \cite{kil07}, who claim
that He  white dwarfs may exist in  the Galaxy as a  product of single
stellar  evolution,  we have  considered  also  this composition.   In
Fig.~\ref{fig:lumtef}  (bottom-left)  we   have  plotted  the  cooling
sequences of  \cite{ser02} for He-core  white dwarfs.  But, as  in the
case of the O/Ne-core models they  do not reproduce a white dwarf with
this    observed   radius.     Moreover,   as    can   be    seen   in
Fig.~\ref{fig:lumtef},  the effective  temperature  and luminosity  of
WD0433$+$270 does  not fall  in the range  of values covered  by these
tracks.  In any case, He-core  white dwarfs cool typically slower than
C/O-core ones because of the  larger heat capacity of He in comparison
with C and O. Although this is partly compensated by the smaller mass,
the net result does not lead to smaller cooling times.

Finally,   we  have  taken   into  account   the  cooling   tracks  of
\cite{pan00b} for  Fe-core white dwarfs,  obtaining a cooling  time of
$\sim$1 Gyr, which is 4 times  smaller than for C/O white dwarfs. This
is because  Fe nuclei  are much heavier  than C  or O and,  hence, the
specific heat per  gram is much smaller.  The  cooling time derived is
in much better agreement with the  age of the Hyades open cluster when
considering an Fe core. It should  be noted that the cooling time of a
white  dwarf  is just  a  lower  limit to  its  total  age, since  the
main-sequence lifetime  of its  progenitor, which is  badly determined
for single white dwarfs, should  also be taken into account.  However,
this value could be relatively  small if the white dwarf progenitor is
massive  ($t_{\rm prog}\ll1$ Gyr).   It is  worth mentioning  that the
uncertainties  in  the  cooling   times  are  derived  only  from  the
observational  parameters and  do not  consider a  possible systematic
contribution from the cooling sequences.

\section{Discussion and Conclusions}

In this work  we have presented new spectroscopic  observations of the
members of  the common proper  motion pair G39$-$27/289.   Using these
observations and the available photometry we have better characterized
these objects, which has helped us to understand their nature.

Considering the  kinematic properties of  the members of the  pair and
the lithium detection in BD$+$26~730,  we favour the scenario in which
the common proper motion pair is  indeed a former member of the Hyades
cluster,  and thus  its members  have a  coeval age  of $\sim$0.6--0.7
Gyr.  Having evaluated  different compositions  for  WD0433$+$270, the
young age inferred from cluster membership is only compatible with the
case of an Fe  core, which  would have an  associated cooling  time of
$\sim$1 Gyr.  The  agreement is not perfect, but  the modelling of the
cooling  evolution  of Fe-core  white  dwarfs  could  still have  some
associated uncertainties.

The existence of white  dwarfs with an Fe-rich core has important
consequences  for the  models  of thermonuclear  explosion of  stellar
degenerate cores.  According to the  theory of stellar  evolution, all
stars that develop an Fe core  experience a collapse to a neutron star
or a  black hole, regardless of  the mass loss  rate assumed. However,
there is  still an  alternative possibility to  avoid this  final fate
that  relies  in the  failure  of  the  thermonuclear explosion  of  a
degenerate  white  dwarf  near   the  Chandrasekhar  limit  (Isern  et
al.~1991).  Our  current  view  of  a thermonuclear  explosion  is  as
follows. Once the thermonuclear  runaway starts in the central regions
of a white  dwarf, the ignition front propagates  outwards and injects
energy  at a  rate that  depends on  the velocity  at which  matter is
effectively burned leading  to the expansion of the  star. At the same
time, electron  captures on the incinerated  matter efficiently remove
energy at a rate determined  by the density causing the contraction of
the  star.  Therefore,  depending  on  the ignition  density  and  the
velocity of  the burning  front, the outcome  can be different.  It is
known  that  He-degenerate cores  always  explode,  that  O/Ne and  Fe
degenerate  cores always  collapse and  that C/O-degenerate  cores can
explode or collapse depending  on the ignition density.  The existence
of Fe-rich white dwarfs would imply the possibility of an intermediate
behaviour between those discussed above in which an Fe remnant is left
after a mild explosion.

A detailed theory explaining the  formation of Fe-core white dwarfs is
still to be developed and up to now the possibility of their existence
has been suggested mainly from observational evidences. Other examples
of   possible  Fe-core   white   dwarfs  have   been   found  in   the
past. \cite{pro98} singled out two objects whose radii and masses were
too small to  fit the typical C/O mass-radius  relation. These authors
analyzed  a sample  of  white dwarfs,  calculated  their masses  using
either  orbital parameters,  gravitational redshifts  or spectroscopic
fits, and  determined their radii independently from  the knowledge of
effective temperatures and  distances.  \cite{pro98} indicated that in
those two cases, a core made  by Fe was the only plausible explanation
that could account for their small radii.

Besides the Fe-core hypothesis,  there is an alternative scenario that
we  cannot rule out.   Although we  think it  is unlikely,  the common
proper  motion pair  might not  be related  in any  way to  the Hyades
cluster. This  would eliminate the  age constraint and thus  relax the
requirement of  an exotic  composition for the  white dwarf  core.  In
this case, the age of  the objects (both WD0433$+$270 and BD$+$26~730)
could  be   of  $\sim$4  Gyr.    This  would  have   some  interesting
consequences to our  current view of star streams  or moving groups in
th Galaxy.  As pointed out by \cite{che99}, the Hyades stream contains
at least  three distinct age groups with  ages up to 2  Gyr. The upper
limit was probably a consequence of the use of A--F stars as kinematic
tracers, which would  naturally have an age cutoff  at about 2--3 Gyr.
With an age of 4  Gyr, WD0433$+$270 and BD$+$26~730 would indicate the
existence  of an  even  older  population in  the  Hyades stream  thus
completely  ruling out any  coevality within  the members.  This would
lend strong support  to the model of a resonant  origin for the Hyades
stream (Famaey et al.~2007). Another implication of such an old age is
the  conflict with  lithium detection,  which would  imply  much lower
destruction rate than expected.

Definitive proof supporting one of  these scenarios will have to await
an  unambiguous   determination  of  the  mass  of   this  object  or,
alternatively, a  more conclusive study on its  evolutionary link with
the Hyades cluster.

\begin{acknowledgements}
We thank  C.~Allende Prieto for  his help during the  observations and
his useful comments in the reduction and analysis of BD$+$26~730. S.~C. 
would like to acknowledge support from MEC through a FPU grant. This
research    was    partially    supported    by   the    MEC    grants
AYA05--08013--C03--01 and 02, by the European Union FEDER funds and by
the AGAUR.
\end{acknowledgements}

\end{document}